\documentclass[prd,aps,twocolumn]{revtex4}

\usepackage[dvips]{graphicx}
\usepackage{amssymb}
\usepackage{amsmath}

\usepackage{color}

\begin{document}

\title{Are symmetric tidal streams possible with long-range dark-matter forces?}

\author{Michael Kesden}

\affiliation{California Institute of Technology,
  MC 130-33, 1200 E. California Blvd., Pasadena, CA 91125}

\date{September 2009}
                            
\begin{abstract}
  The unique dynamics of the tidal disruption of satellite galaxies is
  an extremely sensitive probe of long-range interactions between
  dark-matter particles.  Dark-matter forces that are several percent
  the strength of gravity will lead to order unity changes in the
  ratio of the number of stars in the leading and trailing tidal
  streams of a satellite galaxy.  The approximate symmetry of the
  stellar tidal streams of the Sagittarius dwarf galaxy would thus
  exclude attractive dark-matter forces greater than 10\% the strength
  of gravity which would entirely eliminate the leading stream.
  However, recent simulations suggest that dark-matter forces 100\%
  the strength of gravity could completely strip the stellar component
  of Sagittarius of its dark matter, allowing for the subsequent
  development of symmetric tidal streams.  Here we argue that these
  simulations use inconsistent initial conditions corresponding to
  separate pure stellar and pure dark-matter satellites moving
  independently in the host galaxy's halo, rather than a single
  disrupting composite satellite as had been intended.  A new
  simulation with different initial conditions, in particular a much
  more massive satellite galaxy, might demonstrate a scenario in which
  symmetric tidal streams develop in the presence of large dark-matter
  forces.  This scenario must satisfy several highly restrictive
  criteria described in this paper.
\end{abstract}
\maketitle

\section{Introduction} \label{S:intro}

Understanding the nature of dark matter (DM) remains one of the most
important outstanding problems in cosmology.  If DM is a new
fundamental particle, it may experience new non-gravitational
interactions that would manifest as a violation of the weak
equivalence principle.  Such DM forces should be observationally
constrained on all scales where DM influences structure, from the
subgalactic out to the Hubble radius.  In important early work,
Frieman and Gradwohl \cite{Friedman:1991dj,Gradwohl:1992ue} recognized
that DM forces would violate the equivalence principle, and considered
cosmological constraints on DM forces of the form
\begin{equation}
{\bf F} = -\frac{\beta Gm^2}{r^2} e^{-r/\lambda} \left( 1 +
\frac{r}{\lambda} \right) \hat{{\bf r}} \, ,
\end{equation}
where $G$ is Newton's constant, $m$ is the mass of a DM particle, $r$
is the separation between particles, $\lambda$ is a ``screening
length'' beyond which the force is suppressed, and $\beta$ is a
dimensionless parameter characterizing the relative strength of this
force compared to gravity
\footnote{This definition of $\beta$ is identical to that in
\cite{Nusser:2004qu} and \cite{Peebles:2009th} but differs from that
in my earlier papers \cite{Kesden:2006zb,Kesden:2006vz}.  In my
earlier papers $\beta$ was the charge-to-mass ratio for the DM
particles instead of the ratio between the DM force itself and
gravity.  The DM force is proportional to the square of the charge as
in electromagnetism.  The $\beta$ in this paper is therefore equal to
$\beta^2$ in my previous papers \cite{Kesden:2006zb,Kesden:2006vz}.}.
Observations at the time of Frieman and Gradwohl's work excluded DM
forces with $\beta \gtrsim 1.3$ and $\lambda \gtrsim
\mathcal{O}(1~{\rm Mpc})$.

DM forces received renewed attention after it was realized that they
were generically produced by a coupling to dark energy
\cite{Farrar:2003uw} and could arise naturally in string theory
\cite{Gubser:2004uh,Gubser:2004du}.  They also might agree better with
observations than the standard cosmological constant plus cold dark
matter ($\Lambda$CDM) scenario of structure formation
\cite{Nusser:2004qu}.  Kesden and Kamionkowski
(\cite{Kesden:2006zb,Kesden:2006vz}, hereafter KK06) examined the
effects of DM forces on tidally disrupting satellite galaxies, where
the large difference between the mass $m_{\rm sat}$ of the satellite
galaxy and that of the host galaxy $M_{\rm host}$ establishes a
hierarchy in the relevant energy scales of the problem.  Assuming that
the screening length $\lambda \gtrsim d$, an attractive DM force
increases the satellite's orbital energy per unit mass
\begin{equation}
E_{\rm orb} \simeq \frac{GM_{\rm host}}{d}
\end{equation}
by a factor $\beta f_{\rm DM, host} f_{\rm DM, sat}$, where $d$ is the
distance between the satellite and host galaxies and $f_{\rm DM}$ is
the DM mass fraction in the satellite or host.  A tidally disrupted
star will end up in the leading (trailing) stream if its binding
energy per unit mass is less (greater) than the typical energy
\begin{equation}
E_{\rm bin} \simeq \frac{Gm_{\rm sat}}{r} \, ,
\end{equation}
where $r$ is the radius of the satellite galaxy.  Equating the change
in orbital energy induced by DM forces to the binding energy at the
tidal radius
\begin{equation}
r_{\rm tid} \simeq \left( \frac{m_{\rm sat}}{M_{\rm host}}
\right)^{1/3} d \, ,
\end{equation}
we find that DM forces will produce an order unity change in the ratio
of the number of stars in the leading to those in the trailing tidal
streams when
\begin{equation} \label{E:beta_lim}
\beta \gtrsim \frac{1}{f_{\rm DM, host} f_{\rm DM, sat}}
\left( \frac{m_{\rm sat}}{M_{\rm host}} \right)^{2/3} \, .
\end{equation}

KK06 applied their constraints to the Sagittarius (Sgr) dwarf galaxy,
a Milky Way satellite approximately $d \simeq 17$ kpc from the
Galactic center discovered in 1994 \cite{Ibata:1994fv} after the
publication of Frieman and Gradwohl's work.  The Sgr dwarf has leading
and trailing stellar tidal streams stretching hundreds of degrees
across the sky that have been observed extensively by the Two-Micron
All-Sky Survey (2MASS) \cite{Majewski:2003ux} and Sloan Digital Sky
Survey (SDSS) \cite{Belokurov:2006ms}.  The proposed mass-to-light
ratio $(M/L)_{\rm Sgr} = 14-36 (M/L)_\odot$ of the Sgr dwarf suggests
that it is DM dominated, as is the Milky Way at the perigalactic and
apogalactic distances of the Sgr orbit, about 15 and 60 kpc
respectively \cite{Law:2004ep}.  The current bound mass of the Sgr
dwarf is $(2-5) \times 10^8 M_\odot$, implying $m_{\rm sat}/M_{\rm
host} \simeq 10^{-3}$.  Inserting this into Eq.~(\ref{E:beta_lim})
along with $f_{\rm DM, host} \sim f_{\rm DM, sat} \sim 1$, we find
that DM forces several percent the strength of gravity should induce
order unity changes in the leading-to-trailing ratio of stars in the
Sgr tidal streams.  N-body simulations confirmed that $\beta \gtrsim
0.1$ completely eliminated the leading tidal stream for a wide range
of models of the satellite galaxy, host galaxy, and satellite orbit.
KK06 thus conservatively excluded $\beta \gtrsim 0.1$, as the Sgr
tidal streams are observed to be roughly symmetric.

Keselman, Nusser, and Peebles (\cite{Peebles:2009th}, hereafter KNP09)
have proposed an alternative scenario in which observations of the Sgr
dwarf could be consistent with $\beta \simeq 1$ DM forces.  In this
proposal, the differential acceleration experienced by stars and DM
segregate them from each other well before the first simulated
pericentric passage.  The self-bound stellar remnant of Sgr is free of
DM and thus immune to DM forces.  It proceeds to develop symmetric
tidal streams in the Galactic gravitational potential well.  The
absence of a Sgr DM halo leaves the remaining stars more vulnerable to
tidal heating during pericentric passages.  The resulting heightened
velocity dispersion could be misinterpreted as evidence for DM if the
virial theorem was misapplied to this highly nonvirialized system.
KNP09 argues that just such a misinterpretation could explain the high
observed mass-to-light ratio $(M/L)_{\rm Sgr} = 14-36 (M/L)_\odot$ of
Sgr, which would otherwise be incompatible with the complete
segregation of stars and DM required by $\beta \simeq 1$.

Unfortunately, the initial conditions used in the simulations
presented in KNP09 are not appropriate for describing this alternative
scenario for the formation of the Sgr tidal streams.  These
simulations begin with the satellite galaxy's stellar and DM
distributions moving with the same velocity and centered about the
same point in the host galaxy's halo.  In the absence of DM forces,
two distributions with the same location and velocity are guaranteed
to be gravitationally bound to each other.  However, DM forces change
the orbital energy of the DM particles by an amount $\Delta E \simeq
\beta f_{\rm DM, host} E_{\rm orb}$ that for $\beta = 1$ exceeds the
energy $E_{\rm bin}$ binding these particles to the satellite.  KNP09
notes that DM forces induce ``a nearly complete segregation of the
stars and DM well before the first pericentric passage.''  For these
initial conditions, the stars and DM are already widely separated in
orbital energy before the simulations even begin, though they
initially coincide in phase space.  This point is dramatized by noting
that if these simulations were run backwards in time, by symmetry the
stars and DM would be segregated just as efficiently.  These
simulations actually describe the temporary coincidental convergence
of two different satellites, one purely stellar and the other pure DM,
moving independently on very different orbits in the host galaxy's
halo.

Proper initial conditions would show that this scenario cannot produce
a purely stellar satellite with symmetric tidal streams like the Sgr
dwarf.  The segregation of stars and DM seen in these simulations will
happen on the very first approach of the satellite galaxy to the
Galactic center, long before dynamical friction can put the satellite
onto an orbit resembling that of the current Sgr dwarf.  The satellite
will have fallen in from distances greater than the host galaxy's
virial radius, and having been accelerated by a $\beta = 1$ DM force
will be moving at a velocity well above the gravitational escape
velocity of the system.  Once segregated from their DM halo, the
satellite stars will be moving on a hyperbolic orbit and will be
ejected from the host galaxy after a single pericentric passage.

We have identified three conditions that must be satisfied for the
KNP09 scenario for the formation of the Sgr tidal streams to be
viable:
\begin{enumerate}

\item The DM force must be strong enough that differential acceleration
in the Galactic halo fully segregates the stars and DM prior to the
formation of the tidal streams. \label{C1}

\item The stellar density in the satellite core must be high enough for
it to remain a bound object after being pulled free from its DM halo.
\label{C2}

\item The DM force must not accelerate the satellite beyond the
gravitational escape velocity of the host galaxy before the satellite
stars are pulled from their DM halo. \label{C3}

\end{enumerate}
We will describe our models of the host and satellite galaxies in
Sec.~\ref{S:gal}.  We will then determine whether these models satisfy
the three conditions given above in Sec.~\ref{S:cond}.  We will find
that while $\beta = 1$ DM forces are more than strong enough to
fulfill condition~\ref{C1}, they are too strong to satisfy
condition~\ref{C3}.  Star formation is also unlikely to be efficient
enough in dwarf galaxies to yield stellar densities that satisfy
condition~\ref{C2}.  A summary and some concluding remarks will be
provided in Sec.~\ref{S:disc}.

\section{Galactic Models} \label{S:gal}

\subsection{Host Galaxy} \label{SS:host}

As we shall see in Sec.~\ref{SS:seg}, for $\beta \gtrsim 1$ stars will
become unbound from the satellite DM halo at galactocentric distances
$d$ much greater than the scale lengths of the Galactic bulge or disk.
For convenience, we will therefore model both these components as
point particles at the Galactic center of masses $M_{\rm bulge}$ and
$M_{\rm disk}$ respectively.  Following KNP09, we will adopt values of
$M_{\rm bulge} = 3.4 \times 10^{10} M_\odot$ and $M_{\rm bulge} =
10^{11} M_\odot$ for these quantities.  KNP09 uses the density profile
\begin{equation} \label{E:rhoKNP}
\rho_{\rm DM, host, KNP}(d) = \frac{v_{\rm halo}^2}{2\pi G}
\frac{d^2 + 3b^2}{(d^2 + b^2)^2}
\end{equation}
for the Galactic DM halo, with $v_{\rm halo} = 131.5$ km/s and $b =
12$ kpc.  We will assume spherical symmetry for the both the host and
satellite galaxies, implying that for each component $X$ a mass
\begin{equation} \label{E:rhoM}
M_X(d) = 4\pi \int_{0}^{d} r^2 \rho_X(r)~dr
\end{equation}
is contained within a sphere of radius $d$.  The DM profile of
Eq.~(\ref{E:rhoKNP}) is unsuitable for our purposes because $M_{\rm
DM, host}(d)$ does not asymptote to a finite value at large $d$,
implying that an escape velocity
\begin{equation} \label{E:vesc}
v_{\rm esc}(d) = \left( \frac{2GM(d)}{d} \right)^{1/2}
\end{equation}
cannot be defined.  We will instead use a Navarro-Frenk-White (NFW)
profile \cite{Navarro:1995iw,Navarro:1996gj}
\begin{equation} \label{E:rhoNFW}
\rho_{\rm DM}(d) = \frac{\rho_0}{(r/b)(1 + r/b)^2} \, ,
\end{equation}
with the same scale radius $b = 12$ kpc and the normalization $\rho_0$
chosen to contain the same Galactic DM mass within the initial
apocentric distance $d_{\rm ap} = 80$ kpc simulated in KNP09.  This
implies a virial mass $M_{\rm v, host} = 1.05 \times 10^{12} M_\odot$,
virial radius $d_{\rm v} = 206$ kpc, and concentration $c_{\rm v,
host} \equiv d_{\rm v}/b = 17.2$, all very reasonable values for
describing the Milky Way halo.

\subsection{Satellite Galaxy} \label{SS:sat}

Following the conventions of KNP09, we will model the stellar
component of the satellite galaxy with a modified Hubble profile
\cite{BT}
\begin{equation} \label{E:MHP}
\rho_{\rm \ast, sat}(r) = \rho_1 \left( 1 + \frac{r^2}{r_{c}^2}
\right)^{-3/2} \, ,
\end{equation}
where $r_c = 0.55$ kpc and $\rho_1$ is chosen so that so that a
stellar mass $M_{\rm \ast, sat}(r_t) = 3.0 \times 10^8 M_\odot$ is
contained within the tidal radius $r_t = 1.67$ kpc.  Unlike KNP09 we
will assume that the stellar density drops to zero outside $r_t$ as
our criteria will be applied to the satellite core.  The satellite
galaxy's DM halo will also use the NFW profile of
Eq.~(\ref{E:rhoNFW}), but with a scale radius $r_s = 3.7$ kpc, virial
radius $r_v = 18.65$ kpc, and virial mass $M_{\rm v, sat} = 1.5
\times 10^9 M_\odot$.  We will assume for simplicity that the DM
density vanishes outside $r_v$, as the DM at large radii is largely
irrelevant.  This choice of parameters implies
\begin{equation} \label{E:SFeff}
\frac{M_{\rm \ast, sat}(r_t)}{M_{\rm v, sat}} \simeq
\frac{\Omega_b}{\Omega_{DM}} \simeq 0.2 \, ,
\end{equation}
and therefore that star formation is 100\% efficient in satellite
galaxies.  This assumption is difficult to reconcile with theoretical
arguments suggesting that feedback and reionization suppress star
formation in low-mass halos \cite{Bullock:2000wn}, but we will
consider it as a starting point.

\section{Conditions for Symmetric Tidal Streams} \label{S:cond}

The satellite and host galaxies described in Sec.~\ref{S:gal} must
satisfy three distinct constraints if their interaction is to produce
a tidally disrupting system with symmetric tidal streams in the
presence of a $\beta = 1$ DM force.  We will identify and describe
these three constraints in the three subsections below.

\subsection{DM segregation} \label{SS:seg}

KK06 showed that DM-dominated satellite galaxies will develop
asymmetric tidal streams in the presence of DM forces.  KNP09 argues
that this can be avoided if nearly all of the DM is removed from the
satellite galaxy before its stellar tidal streams begin to form.  As
the satellite galaxy descends deeper into the potential well of the
host galaxy, both stars and DM will be stripped away by conventional
gravitational tidal forces.  This occurs because parts of the
satellite at different distances from the center of the host galaxy
experience different gravitational accelerations.  The weak
equivalence principle implies that at the {\it same} position, all
objects experience the {\it same} gravitational acceleration.
However, DM forces violate the equivalence principle and individual
stars, individual DM particles, and the partly stellar, partly DM
satellite galaxy will experience three different accelerations towards
the host galaxy's DM halo.  In a reference frame freely falling with the
satellite as a whole, the individual stars will experience a relative
acceleration
\begin{equation} \label{E:astarhost}
a_{\rm \ast-host}(d) = \frac{G\beta f_{\rm DM, sat} M_{\rm
DM,host}(d)}{d^2}~,
\end{equation}
while the DM particles experience a relative acceleration
\begin{equation} \label{E:aDMhost}
a_{\rm DM-host}(d) = \frac{G\beta (1 - f_{\rm DM, sat}) M_{\rm
DM,host}(d)}{d^2}~.
\end{equation}
In the absence of DM forces ($\beta = 0$), these relative
accelerations vanish as required by the equivalence principle.
Furthermore, stars will experience no acceleration relative to a
purely stellar ($f_{\rm DM, sat} = 0$) satellite galaxy and DM
particles will similarly experience no acceleration relative a purely
DM ($f_{\rm DM, sat} = 1$) satellite.

Tidal forces, supplemented by these relative accelerations, will
succeed in disrupting the satellite galaxy if they are stronger than
the attractive forces that act to keep stars and DM bound to the
satellite.  For stars this restoring acceleration is simply the
gravitational attraction
\begin{equation} \label{E:astarsat}
a_{\rm \ast-sat}(r) = \frac{G[M_{\rm \ast, sat}(r) + M_{\rm DM,
sat}(r)]}{r^2}
\end{equation}
between the stars and the satellite, while for DM particles this
gravitational restoring force is supplemented by the DM attraction
to the satellite's DM halo
\begin{equation} \label{E:aDMsat}
a_{\rm DM-sat}(r) = \frac{G[M_{\rm \ast, sat}(r) + (1 + \beta) M_{\rm
DM, sat}(r)]}{r^2}~.
\end{equation}

\begin{figure}[t!]
  \begin{center} \includegraphics[width=3.3in]{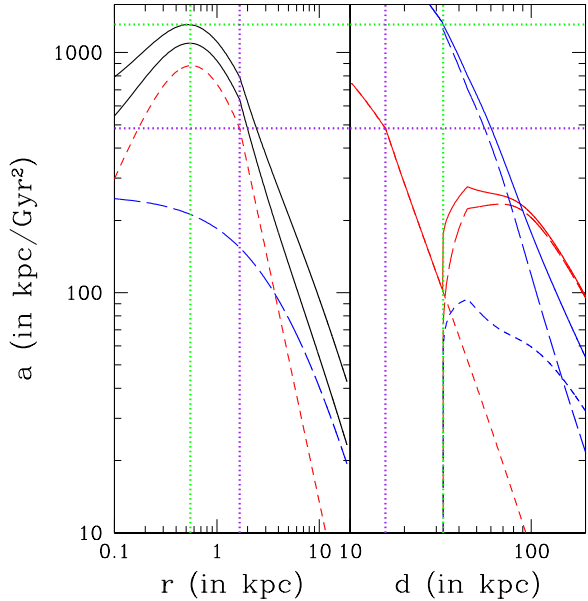}
  \end{center} \caption{The accelerations $a$ experienced by stars and
  DM particles a distance $r$ from the center of the satellite galaxy
  and $d$ from the center of the host galaxy.  {\it Left-hand panel:}
  The lower solid black curve shows the gravitational attraction
  $a_{\rm \ast-sat}$ between stars and the satellite galaxy.  The
  short-dashed red and long-dashed blue curves show the contributions
  of stars and DM respectively to this gravitational attraction.  The
  upper solid black curve shows the acceleration $a_{\rm DM-sat}$
  between DM particles and the satellite; the DM contribution has been
  doubled relative to $a_{\rm \ast-sat}$ for $1 + \beta = 2$.  {\it
  Right-hand panel:} The short-dashed blue curve shows the tidal
  acceleration at the edge of the satellite's DM halo, the long-dashed
  blue curve shows the relative acceleration $a_{\rm DM-host}$, and
  the solid blue curve shows the sum of these two accelerations.  The
  three red curves show the corresponding quantities for the stars.
  The satellite is fully stripped of DM when it first reaches a
  distance $d_{\rm seg} = 32.7$ kpc from the center of the host galaxy
  as shown by the vertical dotted green line in the right-hand panel.
  At this distance, the total relative acceleration experienced by DM
  particles (solid blue curve in right-hand panel) equals the maximum
  of the restoring acceleration (upper solid black curve in left-hand
  panel) which is located near the satellite's stellar core radius
  $r_c = 0.55$ kpc (vertical dotted green line in left-hand panel).
  Tidal disruption of the stars does not begin until $d_{\rm tid} =
  15.7$ kpc (vertical dotted purple line in right-hand panel) when the
  total relative acceleration experienced by stars (solid red curve in
  right-hand panel) equals the restoring acceleration for stars at the
  stellar tidal radius $r_t = 1.67$ kpc (vertical dotted purple line
  in left-hand panel).}  \label{F:acc}
\end{figure}

\begin{figure}[t!]
  \begin{center} \includegraphics[width=3.3in]{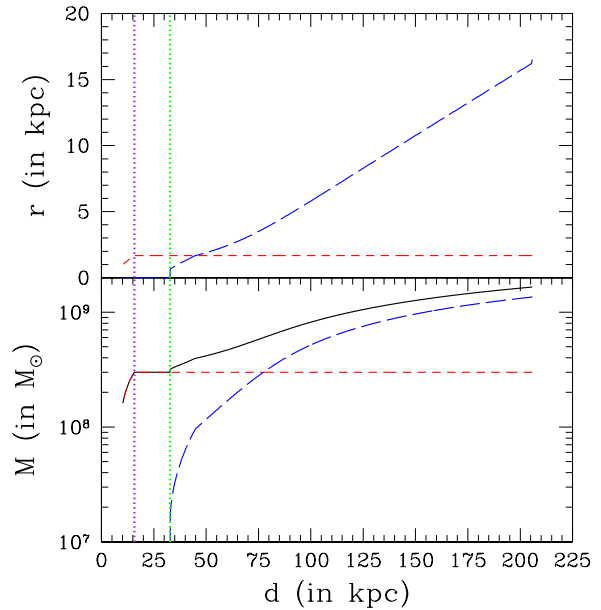}
  \end{center} \caption{{\it Upper panel:} The radius $r$ of
  satellite's DM halo (long-dashed blue curve) and stellar
  distribution (short-dashed red curve) as the satellite gets within a
  distance $d$ of the host galaxy.  {\it Lower panel:} The mass $M$ of
  the satellite galaxy (solid black curve), its DM halo (long-dashed
  blue curve), and stellar distribution (short-dashed red curve) at a
  distance $d$ from the host galaxy.  The vertical dotted green curve
  shows the distance $d_{\rm seg} = 32.7$ kpc at which the stars and
  DM are fully segregated, while the vertical dotted purple curve
  shows the distance $d_{\rm tid} = 15.7$ kpc at which tidal
  disruption of the stellar component begins.} \label{F:disrupt}
\end{figure}

These accelerations are shown in Fig.~\ref{F:acc} for the galactic
models described in Sec.~\ref{S:gal} with a $\beta = 1$ DM force.  The
many curves in this figure are described in its long caption.
Disruption of the satellite's extended DM halo has already begun even
at the host galaxy's virial radius $d_v = 206$ kpc.  This can be seen
by observing that the upper solid black curve $a_{\rm DM-sat}$ in the
left panel intersects the right boundary below where the solid blue
curve $a_{\rm DM-sat}$ in the right panel intersects the right
boundary.  As the satellite falls inwards, the radius $r$ and mass $M$
of its DM halo decrease as shown by the long-dashed blue curves in
Fig.~\ref{F:disrupt}.  Although the tidal forces initially dominate
the disruption, the relative acceleration $a_{\rm DM-host}$ exceeds
the tidal acceleration for $d < 147.6$ kpc where the long-dashed blue
curve crosses the short-dashed blue curve in the right panel of
Fig.~\ref{F:acc}.  Disruption of the satellite's DM halo continues
until $d_{\rm seg} = 32.7$ kpc, when the total relative acceleration
of the DM exceeds the maximum of the restoring acceleration $a_{\rm
DM-sat}$.  At this distance all the remaining DM mass $M_{\rm DM,
sat}(r_c) = 1.6 \times 10^7 M_\odot$ interior to the stellar core
radius $r_c = 0.55$ kpc where this maximum occurs is disrupted {\it en
masse}.  This occurs because $a_{\rm DM-sat}(r)$ is not a
monotonically decreasing function of $r$ as a result of the cored
stellar profile $\rho_{\rm \ast, sat}(r)$ of Eq.~(\ref{E:MHP}).  The
sharp drop in $M_{\rm DM, sat}$ leads to corresponding sharp drops in
$f_{\rm DM, sat}$ and $a_{\rm \ast-host}$ in Eq.~(\ref{E:astarhost}),
explaining the cuspy nature of the total relative acceleration for
stars shown by the solid red curve in the right panel of
Fig.~\ref{F:acc}.  This cored stellar profile is very conducive to the
scenario of KNP09, as a cuspy profile would have had a greater central
restoring acceleration $a_{\rm DM-sat}(0)$ and might have retained DM
until stellar tidal disruption commenced.  The satellite galaxy, now
entirely free of its DM halo, remains undisturbed until gravitational
tidal forces on their own can disrupt it in the standard manner.  This
occurs at a distance $d_{\rm tid} = 15.7$ kpc from the host galaxy
where the tidal forces exceed the restoring acceleration $a_{\rm
\ast-sat}(r_t)$ at the satellite's tidal radius $r_t = 1.67$ kpc.

Simulation S2 of KNP09 was designed to illustrate how symmetric tidal
streams resembling those of the Sgr dwarf could come about even in the
presence of $\beta = 1$ DM forces.  It begins with the satellite at an
apocentric distance of $d_{\rm ap} = 84.8$ kpc.  Although the
pericentric distance is not provided explicitly, if the streams are to
resemble those of Sgr it must be close to Sgr's estimated pericentric
distance of $d_{\rm pe} \simeq 14$ kpc \cite{Law:2004ep}.  For this
choice of an initial orbit we have
\begin{equation} \label{E:dhier}
d_{\rm pe} < d_{\rm tid} < d_{\rm seg} < d_{\rm ap}
\end{equation}
which demonstrates one of the concerns with this simulation.  Although
the satellite begins the simulation with its DM halo intact out to the
virial radius $r_v$, this extended halo should have been disrupted at
much greater distances from the host galaxy.  Since $d_{\rm pe} <
d_{\rm seg}$, the satellite should not have {\it any} DM halo at all.
These initial conditions represent a temporary association of stars
and DM moving independently in the Galactic halo, like ships passing
in the night.  To properly simulate this scenario, the satellite's
orbit must evolve from that on which the DM was fully segregated
($d_{\rm pe} = d_{\rm seg}$) to the current orbit needed to produce
the observed stellar tidal streams.

Dynamical friction drives this orbital evolution, but was neglected in
KNP09 since the streams develop on a few orbital times $t_{\rm orb}$
and according to \cite{BT} the dynamical-friction time
\begin{equation} \label{E:DFt}
t_{\rm DF} \sim \left( \frac{M_{\rm host}}{m_{\rm sat}} \right)
t_{\rm orb} \simeq 10^3 t_{\rm orb}
\end{equation}
is much longer than this for the Sgr-Milky Way system.  In response to
this paper, KNP09 performed a new simulation with a ``live'' DM halo
for the host galaxy that allowed for dynamical friction and a more
massive satellite $m_{\rm sat}/M_{\rm host} \simeq 0.1$ that would
reduce the dynamical-friction time $t_{\rm DF}$.  The satellite also
began on an orbit with a greater apocentric distance $d_{\rm ap} =
250$ kpc to allow for orbital evolution.  A detailed analysis of this
new simulation is beyond the scope of this paper, but it should be
examined whether this live halo can adequately describe the dynamical
friction given that the final satellite mass is $m_{\rm sat} \lesssim
4 \times 10^8 M_\odot$ and the individual N-body particles of the host
galaxy's DM halo have masses of $10^8 M_\odot$.

The segregation distance $d_{\rm seg} = 32.7$ kpc for the Sgr dwarf is
comparable to the current distances of the newly discovered
ultrafaint SDSS dwarf galaxies Segue 1, Ursa Major II, Willman 1, and
Coma, all of which lie within 50 kpc of the Galactic center.  These
new ultrafaint dwarf galaxies have a wide range of luminosities
$(10^2 \lesssim L/L_\odot \lesssim 10^7)$, yet share a common central
density $M(r = 0.3~{\rm kpc}) \simeq 10^7 M_\odot$
\cite{Strigari:2008ib}.  Simulations show that tidal heating cannot
inflate the velocity dispersions of these satellites enough to
eliminate the need for massive DM halos \cite{Fellhauer:2008ia}.  The
satellites less luminous than Sgr will thus have higher DM fractions
$f_{\rm DM, sat}$ and correspondingly higher relative accelerations
$a_{\rm \ast-host}$ for stars according to Eq.~(\ref{E:astarhost}).
The common central density however implies that the restoring
acceleration $a_{\rm \ast-sat}$ for the stars will remain largely
independent of satellite luminosity.  The stars should therefore be
fully stripped from their DM halos at greater distances than $d_{\rm
seg} = 32.7$ kpc for the more massive Sgr dwarf.  This is not observed
for Segue 1, the closest of the new dwarf satellites, whose
galactocentric distance of 28 kpc \cite{Geha:2008zr} is well within
its predicted $d_{\rm seg}$ for $\beta = 1$.  Although the ultrafaint
dwarf satellites appear to place even tighter constraints on DM
forces, if we restrict our attention to Sgr we see that a $\beta = 1$
DM force is strong enough to segregate stars and DM in Sgr at
distances $d_{\rm seg} > d_{\rm tid}$ before the stellar tidal streams
begin to form.  The first of our three criteria for Sgr to form
symmetric tidal streams as proposed in KNP09 is thus satisfied.

\subsection{Intact Stellar Core} \label{SS:surv}

Even if DM forces are strong enough to fully segregate a satellite
galaxy's stars from their DM halo, the segregated stars may not
survive as a self-bound object after being pulled free.  The survival
of an intact stellar core is essential to the scenario proposed in
KNP09, as it is this remnant that subsequently develops the symmetric
tidal streams observed today.  The stellar core is threatened with
destruction at three distinct stages:
\begin{enumerate}

\item The satellite galaxy may be stripped of all its stars before
the stars and DM are fully segregated.

\item The stellar component of the satellite may never have been
self-bound, but merely held together by the gravity of the satellite's
DM halo.

\item Even if the stars are initially self-bound, they may be tidally
disrupted by the satellite's DM halo as it is segregated.

\end{enumerate}
We will explore each of these threats in turn in the remainder of this
subsection.

\subsubsection{Competitive Segregation} \label{SSS:comp}

We saw in the previous subsection that stars will be stripped from the
satellite when the sum of the gravitational tidal acceleration and the
relative acceleration $a_{\rm \ast-host}$ exceed the restoring
acceleration $a_{\rm \ast-sat}$, while DM particles will be stripped
when the same inequality is reached between $a_{\rm DM-host}$ and
$a_{\rm DM-sat}$.  This establishes a competition between stars and
DM: if the DM profile is sufficiently cuspy or the satellite is
sufficiently DM-dominated, there may not be a stellar remnant left
when the satellite finally reaches $d_{\rm seg}$.  Returning to the
accelerations plotted in Fig.~\ref{F:acc}, we see that at large $d$
the total relative acceleration for stars shown by the solid red curve
in the right-hand panel was above the total relative acceleration for
DM shown by the solid blue curve.  If the satellite's DM halo had been
more massive or less extended compared to the stellar component,
$f_{\rm DM, sat}$ would have remained close to unity and the total
relative acceleration for stars could have exceeded the maximum of the
lower black curve in the left-hand panel before this happened for the
DM at $d_{\rm seg}$.  In this case, all of the stars exterior to this
maximum located at $r \simeq r_c$ would have been disrupted into
asymmetric tidal streams.  A small stellar remnant would remain
because all of the stars interior to $r_c$ would have been stripped
{\it en masse} as was the case for DM in our model of the Sgr dwarf in
the previous subsection.  If the restoring acceleration $a_{\rm
\ast-sat}$ had been monotonically decreasing no bound stellar remnant
would survive.

Though we have estimates of the total mass and mass-to-light ratio of
Sgr, the stellar and DM density profiles themselves are very
uncertain.  It is worth exploring more generally which profiles will
allow an intact stellar core to be segregated from its DM halo.  A
general DM density profile will scale as $\rho_{\rm DM}(r) \propto
r^{-\alpha}$ at small $r$; the NFW profile of Eq.~(\ref{E:rhoNFW}) has
$\alpha = 1$.  The profile $\alpha = 1$ is special in that the
resulting acceleration $a_{\rm DM}(r) = GM_{\rm DM}(r)/r^2 \propto
r^{1-\alpha}$ will be independent of $r$.  DM halos with $\alpha > 1$
yield accelerations that diverge at $r = 0$ and monotonically decrease
with $r$.  Density profiles with $\alpha < 1$ will produce
accelerations that increase with $r$, before ultimately turning over
at larger $r$ since the halo must have finite mass.

Of these three possibilities, the $\alpha = 1$ profile is the most
conducive to leaving a bound stellar remnant behind.  DM halos with
$\alpha > 1$ will hold onto their innermost stars so tightly that
$d_{\rm seg} = 0$ will always be less than $d_{\rm tid}$ at which
stellar tidal disruption begins.  Simulation S3 of KNP09, though still
employing an NFW profile for the satellite halo, has an enhanced
central concentration that mimics this possibility.  When this
simulation is run, the majority of stars are indeed stripped from the
satellite halo while the inner stellar core remains bound.  Satellites
with $\alpha < 1$ will have accelerations that vanish at $r = 0$, so
even small differential accelerations $a_{\rm dif}$ will displace the
stars from the center of their DM halo.  As the satellite falls
inwards, $a_{\rm \ast-host}$ increases and the stars march up the DM
potential well to ever larger accelerations $a_{\rm \ast-sat}$ before
breaking free from the satellite when they reach the global maximum.
However, because the satellite stellar distribution has a finite
spatial extent, its outer edges will spill over the global maximum
before the center arrives there.  This may cause additional stars to
be stripped from the satellite before the eventual bound stellar
remnant itself breaks free.

If $\alpha$ was truly unity for an NFW profile, not just for $r \ll
b$, the acceleration $a_{\rm DM}(r)$ depicted by the long-dashed blue
curve in the left-hand panel of Fig.~\ref{F:acc} would be flat.  If
star formation was much less efficient as seems to be the case for
less massive dwarf satellites, the DM contribution would dominate in
$a_{\rm \ast-sat}$ and the lower solid black curve would be flat as
well.  In this case, the total relative acceleration would rise above
$a_{\rm \ast-sat}(r)$ simultaneously over a wide range in $r$.  The
stars would slide out of their DM halo like ``an egg sliding off a
frying pan'' \cite{Peebles}, and the stars would be more likely to
remain self-bound.  In the competition to remain bound to the
satellite galaxy, stars are more likely to defeat DM if stellar
densities are high and centrally concentrated, while the DM density
profile's logarithmic slope scales as $\alpha = 1$ over a wide range.
 
\subsubsection{Self-boundedness} \label{SSS:sb}

After being segregated from its DM halo, the stellar remnant may no
longer have enough self-gravity to remain bound.  Unbound stars in the
host galaxy's halo would develop into an orphan stream lacking a clear
progenitor.  One such system, apply dubbed the ``Orphan Stream'', was
discovered serendipitously by SDSS \cite{Belokurov:2006kc}.  This
inspired simulation S4 in KNP09 which demonstrates just how readily
such systems can be produced following the segregation of stars and DM
by large DM forces.  The possibility of making an orphan stream is a
problem for efforts to explain the Sgr dwarf, which has a well defined
bound core identified by a cusp in the M-giant density distribution
spatially coincident with the globular cluster M54
\cite{Majewski:2003ux}.  This bound core serves as a marker to distinguish
the leading and trailing tidal streams of Sgr, essential to the
proposal of KK06 to constrain DM forces by comparing the stars in each
group.

Whether the satellite stars remain bound after their DM halo is
removed depends on their distribution function $f_{\ast, {\rm
sat}}(\vec{r}, \vec{v})$.  If the stars are segregated from the DM on
a dynamical time that is shorter than the relaxation time on which
their individual energies change, one can use an impulse approximation
to determine which stars remain bound.  Without the negative
contribution of the DM halo's gravitational potential, some portions
of phase space that were previously bound will now have positive
energies.  The integral of $f_{\ast, {\rm sat}}(\vec{r}, \vec{v})$
over this portion of phase space will determine the initial mass loss.
Without these stars the satellite's potential well will become even
shallower, and the stars will rearrange themselves with further mass
loss likely.  Unfortunately, many different distribution functions
will yield the density profile of Eq.~(\ref{E:MHP}).  Distinguishing
between them requires knowledge of the anisotropy of the satellite's
velocity dispersion, which is even more poorly constrained
observationally than the density profile itself.

\begin{figure}[t!]
  \begin{center} \includegraphics[width=3.3in]{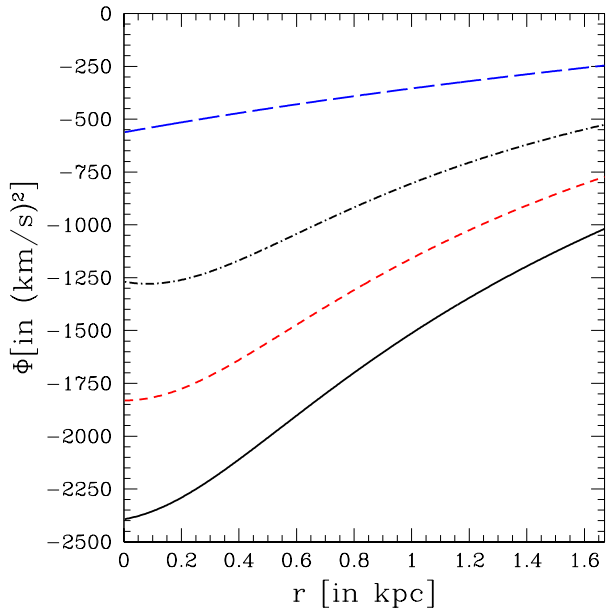}
  \end{center} \caption{Gravitational potentials $\Phi$ as a function
  of the distance $r$ from the center of the satellite galaxy.  The
  short-dashed red curve is the potential $\Phi_\ast$ from stars, the
  long-dashed blue curve is the potential $\Phi_{{\rm DM}<r_t}$ from
  DM interior to the tidal radius $r_t$, and the solid black curve is
  their sum.  The dot-dashed black curve is their difference
  $\Phi_\ast - \Phi_{{\rm DM}}$ whose stellar-mass-weighted average,
  appearing in Eq.~(\ref{E:Ef}), determines whether the satellite
  stars remain bound after the DM halo has been removed.}
  \label{F:pot}
\end{figure}

One can crudely estimate whether the satellite stars remain bound by
applying the virial theorem to the initial system to determine its
kinetic energy
\begin{subequations} \label{E:kin}
  \begin{eqnarray}
	K &=& -\frac{1}{2} W_i \\
	  &=& -\frac{1}{4} \int \rho_{\ast}(r)
	[\Phi_{\ast}(r) + \Phi_{{\rm DM} < r_t}(r)]
	~d^3\vec{r}~.	
  \end{eqnarray}
\end{subequations}
Here $W_i$ is the initial gravitational potential energy of the stars,
$\Phi_{\ast}(r)$ is the gravitational potential of the satellite
stars, and $\Phi_{{\rm DM} < r_t}(r)$ is the gravitational potential
sourced by DM located interior to $r_t$.  Spherical symmetry implies
that only the mass interior to $r_t$ can accelerate the stars and thus
contribute to their kinetic energy $K$.  If the DM halo is removed
quickly enough, this kinetic energy $K$ will remain unchanged while
the final potential energy will be
\begin{equation} \label{E:potF}
W_f = \frac{1}{2} \int \rho_{\ast}(r) \Phi_{\ast}(r)~d^3\vec{r}
\end{equation}
now that the DM halo has been removed.  The total final energy $E_f$
can be found by summing Eqs.~(\ref{E:kin}) and (\ref{E:potF})
\begin{subequations} \label{E:Ef}
  \begin{eqnarray}
	E_f &=& K + W_f \\ \label{E:potdif}
	    &=& \frac{1}{4} \int \rho_{\ast}(r)
	[\Phi_{\ast}(r) - \Phi_{{\rm DM} < r_t}(r)]
	~d^3\vec{r} \, .
  \end{eqnarray}
\end{subequations}
If $E_f > 0$, the satellite as a whole will be unbound, though a bound
remnant may still remain if the unbound stars carry away enough
energy.  Figure~\ref{F:pot} shows $\Phi_{\ast}$ and $\Phi_{{\rm DM} <
r_t}$ for the galactic models described in Sec.~\ref{S:gal}.  The
dot-dashed black curve shows that their difference, appearing in the
integrand of Eq.~(\ref{E:Ef}), is negative for $r \leq r_t$.  Since
$\rho_{\ast}(r)$ is positive definite, $E_f$ must be negative implying
that the stellar remnant after DM segregation remains self-bound.  If
star formation had been less efficient or the DM profile was cuspier,
we could have had $\Phi_{\ast} > \Phi_{{\rm DM} < r_t}$ for some $r$
which might have driven $E_f$ positive.  Only star formation
efficiencies approaching the theoretical maximum of
Eq.~(\ref{E:SFeff}) allow the possibility of leaving a bound stellar
remnant behind after the stars are segregated from their DM halo.

\subsubsection{Tidal Disruption} \label{SSS:td}

Even if the stellar remnant of the satellite is self-bound when
initially displaced from its DM halo, it may be tidally disrupted by
the satellite DM halo's gravity before it escapes beyond the satellite
virial radius $r_v$.  This process is in some ways the reverse of
traditional tidal disruption, in that the stars begin at the center $r
= 0$ of the satellite DM halo and are pulled to larger $r$ by the
relative acceleration, rather than beginning at a large distance $d$
from the host and falling inwards towards $d = 0$.  The tidal
interactions between comparable-mass systems like the satellite
stellar and DM distributions are extremely difficult to model
analytically, necessitating the N-body simulations used by both KK06
and KNP09.  The best we can hope to do without such simulations is to
compare the restoring acceleration at the edge of the stellar remnant,
given by the stellar contribution to $a_{\rm \ast-sat}(r_t)$, to the
tidal acceleration sourced by the satellite's DM halo given by the DM
contribution to $|a_{\rm \ast-sat}(r) - a_{\rm \ast-sat}(r + r_t)|$.
The left-hand panel of Fig.~\ref{F:acc} shows that the stellar
contribution to $a_{\rm \ast-sat}$ (short-dashed red curve) in the
range $0 \leq r \leq r_t$ is greater than the DM contribution
(long-dashed blue curve) over the entire range $0 \leq r \leq \infty$.
This implies that the satellite DM halo cannot disrupt the stellar
remnant during the segregation of the two components.  As for the
other two constraints considered in this section, this might not have
been true for a cuspier DM profile or less massive stellar component
of the satellite galaxy.

\subsection{Satellite Bound to Host Galaxy} \label{SS:bound}

\begin{figure}[t!]
  \begin{center} \includegraphics[width=3.3in]{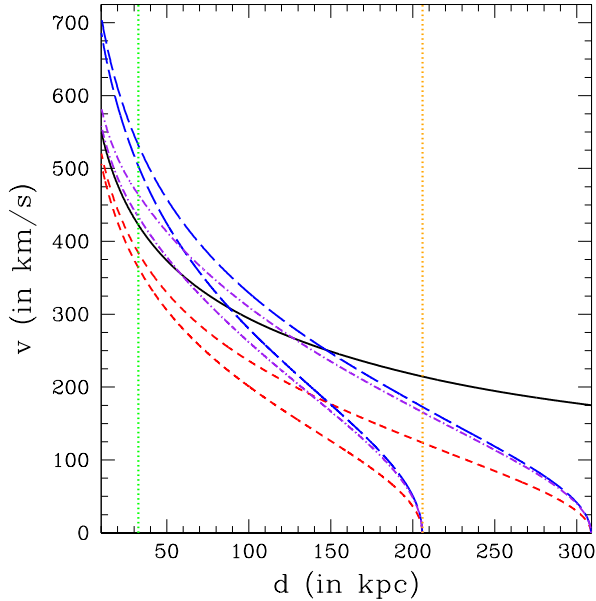}
  \end{center} \caption{Velocities $v$ of infalling stars, DM
  particles, and the satellite galaxy as a function of their distance
  $d$ from the host galaxy.  These velocities are for radial infalls
  beginning at rest from turnaround distances $d_{\rm ta}$ equal to
  the host galaxy's virial radius $d_v = 206$ kpc and $1.5 d_v$.  The
  short-dashed red curves correspond to freely falling stars, the
  long-dashed blue curves to DM particles, and the dot-dashed purple
  curves to the satellite galaxy which is partly stellar and partly
  DM.  The solid black curve shows the gravitational escape velocity
  $v_{\rm esc}$.  The dotted vertical orange line shows the host
  virial radius $d_v$, while the dotted vertical green line shows the
  segregation distance $d_{\rm seg}$.}\label{F:v}
\end{figure}

The final condition that must be satisfied for the KNP09 scenario for
the formation of symmetric tidal streams is that the satellite stars
remain bound to the host galaxy after they are segregated from their
DM halo.  This is a significant concern, as even in the absence of DM
forces the energy
\begin{equation} \label{E:Etid}
E_{\rm tid} \simeq \frac{\partial \Phi_{\rm host}}{\partial d} r_{\rm tid}
\simeq \left( \frac{m_{\rm sat}}{M_{\rm host}} \right)^{1/3} E_{\rm orb}
\end{equation}
gained during tidal disruption can sometimes unbind stars in the
trailing tidal stream from the host galaxy.  This energy $E_{\rm tid}$
is responsible for raising extended tidal tails in interacting
galaxies, and has even been proposed as an explanation for the large
fraction of hypervelocity stars with common travel times clustered in
the direction of the constellation Leo \cite{Abadi:2008ia}.  The
satellite galaxies in simulations S1 and S2 of KNP09 begin $\sim 80$
kpc from the host galaxy with velocities of 80 km/s, much less than
the escape velocity $v_{\rm esc} = 333$ km/s at this distance.
Dynamical friction over several orbits is assumed to have placed the
satellite on this orbit, similar to that of the Sgr dwarf, before
segregation occurs and the tidal streams begin to develop.  However,
we saw in Sec.~\ref{SS:seg} that the stars will be segregated from
their DM halo the first time the satellite gets with a distance
$d_{\rm seg}$ of the host galaxy.  The satellite will be traveling at
a greater velocity on this first approach and therefore its stars are
more likely to become unbound from the host galaxy.

The secondary infall of subsequent material onto an already collapsed
structure has been studied extensively in a cosmological context
\cite{Gott:1975,Fillmore:1984wk,Bertschinger:1985pd}.  At early times
this material recedes with the Hubble expansion, before reaching a
maximum turnaround distance $d_{\rm ta}$ and falling back onto its
host galaxy.  Models of secondary infall can be used to predict the
orbits of satellite galaxies \cite{White:1992}, and comparison with
observations suggest $d_{\rm ta} \simeq 1$ Mpc for the Milky Way
\cite{Zaritsky:1993af}.  Surviving satellites like the Sgr dwarf 
are biased towards large turnaround distances $d_{\rm ta}$ compared to
disrupted satellites that form the Galactic stellar halo
\cite{Sales:2007hp}.

In Fig.~\ref{F:v} we show the velocities acquired by stars, DM
particles, and the satellite as they freely fall towards the host
galaxy on radial orbits from the turnaround distance.  These estimated
velocities are conservative, as the tangential velocities must be
added in quadrature to determine the true total velocity.  The
velocities are calculated by assuming conservation of energy
\begin{equation} \label{E:Econ}
\frac{1}{2} v_{\rm ff}^{2}(d) = \int_{d}^{d_{\rm ta}} a_{\rm ff}(x)~dx~,
\end{equation}
where the different free-fall accelerations $a_{\rm ff}$ for the
stars, DM particles, and satellite are
\begin{subequations} \label{E:aff}
  \begin{eqnarray} \label{E:affstar}
	a_{\rm \ast-ff}(d) &=& \frac{G[M_{\rm \ast, host} +
	M_{\rm DM, host}(d)]}{d^2} \\ \label{E:affDM}
	a_{\rm DM-ff}(d) &=& \frac{G[M_{\rm \ast, host} +
	(1 + \beta)M_{\rm DM, host}(d)]}{d^2} \\ \label{E:affsat}
	a_{\rm sat-ff}(d) &=& \frac{G[M_{\rm \ast, host} +
	(1 + \beta f_{\rm DM, sat})M_{\rm DM, host}(d)]}{d^2}~.
  \end{eqnarray}
\end{subequations}
The DM fraction $f_{\rm DM, sat}$ in Eq.~(\ref{E:affsat}) is
calculated self-consistently as the ratio of the satellite's DM and
total masses, given respectively by the long-dashed blue and solid
black curves in the bottom panel of Fig.~\ref{F:disrupt}.  Note that
the difference $|a_{\rm sat-ff} - a_{\rm \ast-ff}|$ of these free-fall
accelerations gives the relative acceleration $a_{\rm \ast-host}$
appearing in Eq.~(\ref{E:astarhost}).  The difference $|a_{\rm DM-ff}
- a_{\rm sat-ff}|$ similarly gives the relative acceleration $a_{\rm
DM-host}$ in Eq.~(\ref{E:aDMhost}).

The attractive DM force proportional to $\beta$ in
Eqs.~(\ref{E:affDM}) and (\ref{E:affsat}) accelerates the DM particles
and satellite to higher velocities than the stars can attain by
gravity alone.  This can be seen in Fig.~\ref{F:v} where the DM
velocities $v_{\rm DM-ff}$ given by the long-dashed blue curves and
the satellite velocities $v_{\rm sat-ff}$ given by the dot-dashed
purple curves exceed the stellar velocities $v_{\rm \ast-ff}$ given by
the short-dashed red curves.  The satellite is DM dominated ($f_{\rm
DM, sat} \simeq 1$) at large distances, implying that $v_{\rm DM-ff}$
and $v_{\rm sat-ff}$ are initially close together.  As the satellite
falls inwards and loses its DM halo, it becomes dominated by stars
($f_{\rm DM, sat} \simeq 0$)and $v_{\rm sat-ff}$ approaches $v_{\rm
\ast-ff}$.  Crucially however, $v_{\rm sat-ff}$ remains above $v_{\rm
\ast-ff}$; for the models of Sec.~\ref{S:gal} it is even above
the gravitational escape velocity $v_{\rm esc}$ at $d_{\rm seg}$.  The
satellite is accelerated by the DM force on its DM halo while falling
towards $d_{\rm seg}$, but once this halo has been removed the DM
force cannot help keep the stellar remnant bound to the host galaxy
after the first pericenter passage.  It should therefore be ejected
from the host galaxy, not evolve onto the tightly bound orbit needed to
produce stellar tidal streams resembling those of the Sgr dwarf as
predicted by KNP09.

Dynamical friction, described by the Chandrasekhar formula
\cite{Chandrasekhar:1943ys,BT}
\begin{equation} \label{E:DF}
\frac{d{\bf v}_{\rm sat}}{dt} \simeq -\frac{G^2 m_{\rm sat} \rho_{\rm host}}
{v_{\rm sat}^{2}} \hat{{\bf v}}_{\rm sat}~,
\end{equation}
might help keep the stellar remnant bound by reducing its velocity
below the escape velocity.  However Eq.~(\ref{E:DFt}), readily derived
from the Chandrasekhar formula, suggests that dynamical friction will
not be significant on an orbital time for a binary system with a mass
ratio as small as that between the Sgr dwarf and Milky Way.
Increasing the mass of the satellite $m_{\rm sat}$ can increase the
effect of dynamical friction, but introduces problems as well.  A
higher satellite mass will increase the tidal energy $E_{\rm tid}$ in
Eq.~(\ref{E:Etid}), which on its own can unbind tidal debris even in
the absence of DM forces.  The restoring acceleration $a_{\rm DM-sat}$
will also increase with the satellite mass, reducing $d_{\rm seg}$ and
increasing the distance over which the DM force has the chance to
accelerate the satellite beyond the escape velocity according to
Eq.~(\ref{E:Econ}).  An N-body simulation will be required to
determine which of these effects dominates for a particular model.

Just such a simulation was performed in the latest version of KNP09 in
response to the concerns expressed in this paper.  This simulation has
precisely the features necessary to avoid ejecting the stellar remnant
of Sgr from the Galactic halo.  The extraordinarily massive Sgr dwarf
$(M_{\ast, {\rm sat}} = 4 \times 10^{10} M_\odot, M_{\rm DM, sat} = 2
\times 10^{11} M_\odot)$ experiences strong dynamical friction
according to Eq.~(\ref{E:DF}), and its large tangential velocity
$v_{\rm sat} = v_{\rm esc}(d = 250~{\rm kpc}) = 185$ km/s allows time
for this dynamical friction to remove orbital energy before the first
close approach to the Galactic center.  This simulation demonstrates a
possible scenario in which the Sgr dwarf could arrive on its present
orbit in the presence of $\beta
\simeq 1$ DM forces, however several questions remain.  The Sgr dwarf
loses more than 99\% of its stellar mass during the course of the
simulation.  Is this large a contribution to the Galactic stellar halo
consistent with observations?  Although the Galactic DM halo has now
been made of ``live'' particles to allow for dynamical friction, the
Galactic disk is still modeled by a static potential.  Can the
Galactic disk ($M_{\rm disk} = (4.5 \pm 0.5)
\times 10^{10} M_\odot$ \cite{BT}) survive an encounter with such a
massive satellite, particularly in the presence of DM forces?  The
simulated Galactic DM halo consists of $2 \times 10^4$ particles, each
of mass $10^8 M_\odot$.  Is this sufficient to resolve the dynamical
friction on the Sgr stellar core, which also has a mass $M_{\ast, {\rm
sat}} \simeq 10^8 M_\odot$ after losing 99\% of its mass to tidal
disruption?  While this new simulation is intriguing, further
simulations and comparisons to observations are needed to confirm the
validity of the KNP09 scenario.

An alternative possibility to avoid ejecting the stellar remnant of
the satellite is if it was accreted by the host galaxy as part of a
larger group of satellites.  Many of the Milky Way's brightest
satellites appear to lie in a disk, as would naturally occur if these
satellites were initially members of a group that was subsequently
tidally disrupted in the Galactic halo \cite{Li:2007mf}.  Lake and
D'Onghia \cite{Lake:2008zt} specifically proposed that the Magellanic
clouds formed the core of such a group that also included seven other
Milky Way satellite galaxies including the Sgr dwarf.  However, others
have argued that this hypothesized group would have to be more tightly
bound than observed dwarf-galaxy associations, and that independent
accretion from a common filamentary structure could also explain the
observed disk of satellites \cite{Metz:2009ys}.  If the Sgr dwarf was
accreted as the more massive member of a satellite-galaxy binary, it
could lose energy to its lighter companion as the binary was tidally
disrupted in the Galactic halo \cite{Sales:2007hr}.  Such a ``cosmic
m\'{e}nage \`{a} trois'' might conceivably leave the Sgr dwarf with a
small enough orbital energy that its stellar remnant might remain
bound to the Milky Way despite the additional acceleration due to
$\beta = 1$ DM forces.

\section{Discussion} \label{S:disc}

KNP09 performed a series of simulations that suggested that the
symmetric tidal streams of the Sgr dwarf galaxy might be consistent
with DM forces comparable in strength to gravity, in contradiction
with the claims of KK06.  These simulations used static potentials for
the Galactic bulge, disk, and DM halo.  Such static potentials save
computational resources and avoid artificial heating of the tidal
streams, making them essential to attempts like those in
\cite{Law:2004ep} to compare the observed and simulated velocity
dispersions in the tidal streams.  However, static potentials do not
allow for dynamical friction and therefore the satellite galaxy must
begin on its current orbit to reproduce the observed tidal streams.
Such a cheat was acceptable in \cite{Law:2004ep} and KK06, but fails
in the scenario of KNP09 where the segregation of stars and DM should
occur on the satellite's {\it first} approach within a distance
$d_{\rm seg}$ of the host galaxy.  This first approach will likely
occur on a very different orbit from the current one with apocentric
distance $d_{\rm ap} \simeq 80$ kpc needed to produce the observed Sgr
tidal streams.  The initial conditions used in the first series of
simulations in KNP09, with the satellite stars and DM coincident at
$d_{\rm ap}$ with the same velocity, actually correspond to already
unbound stellar and DM distributions moving on widely separated orbits
in the Galactic halo that happened to converge at $t = 0$ when the
simulations begin.

We have considered the first approach of a satellite galaxy to the
Galactic center in the presence of large DM forces, and have
identified three preconditions for the creation of a purely stellar
satellite that can go on to form symmetric tidal streams:
\begin{enumerate}

\item The DM force must be strong enough that differential acceleration
in the Galactic halo fully segregates the stars and DM prior to the
formation of the stellar tidal streams.

\item The stellar density in the satellite core must be high enough for
it to remain a bound object after being pulled free from its DM halo.

\item The DM force must not accelerate the satellite beyond the
gravitational escape velocity of the host galaxy before the satellite
stars are pulled from their DM halo.

\end{enumerate}
The extremely large mass-to-light ratios observed even in the cores of
Milky Way satellites are difficult to reconcile with condition 2, and
such efficient star formation contradicts our theoretical
understanding of the vulnerability of these satellites to reionization
and stellar feedback.  DM forces strong enough to satisfy condition 1
are likely to be too strong to satisfy condition 3 unless a further
epicycle is added to give the Sgr dwarf an initial orbital energy much
less than that expected for a newly accreted satellite galaxy.  While
the scenario proposed in KNP09 for symmetric Sgr tidal streams is
perhaps possible for certain initial orbits and galactic models, it
requires that several restrictive assumptions be satisfied.  Further
simulations are needed that self-consistently capture both the
segregation of stars and DM and the subsequent development of tidal
streams.

Future observations of other Milky Way satellites will also severely
constrain the existence of $\beta \simeq 1$ DM forces.  Once their
proper motions have been measured, the pericenters of their orbits can
be determined.  If these pericenters are too close to the Galactic
center, the DM forces they will experience during pericentric passages
will be inconsistent with retaining their DM halos in seeming
contradiction to the high observed mass-to-light ratios.  More
sophisticated simulations and observations by upcoming astrometry
missions like SIM Lite thus provide a clear path towards closing
remaining loopholes in constraints on DM forces.

\section*{Acknowledgements}  

We would like to thank Andrew Benson, Marc Kamionkowski, Ariel
Keselman, and Jim Peebles for useful discussions.  An anonymous
referee also provided useful comments.  This research was supported by
NASA ATP/BEFS grant NNX07AH06G (PI: Phinney).

\end{document}